\documentclass[5p,sort&compress,10pt]{elsarticle}

\usepackage[]{graphicx}
\usepackage[cmex10]{amsmath}
\usepackage{amssymb}
\usepackage{array}
\usepackage{epstopdf}

\journal{Journal of Neuroscience Methods}

\begin{document}

\begin{frontmatter}
\title{The Complex Hierarchical Topology of EEG Functional Connectivity}

\author[label1,label2]{Keith~Smith$^{*,}$}	

\author[label1]{Javier~Escudero}%
	\address[label1]{Institute for Digital Communications, School of Engineering, University of Edinburgh, West Mains Rd, Edinburgh, EH9 3FB, UK, e-mail: k.smith@ed.ac.uk, javier.escudero@ed.ac.uk}%
	\address[label2]{Alzheimer Scotland Dementia Research Centre, University of Edinburgh, 7 George Square, Edinburgh, EH8 9JZ, UK}%


\begin{abstract}

\textit{Background}: Understanding the complex hierarchical topology of functional brain networks is a key aspect of functional connectivity research. Such topics are obscured by the widespread use of sparse binary network models which are fundamentally different to the complete weighted networks derived from functional connectivity.

\textit{New Methods}: We introduce two techniques to probe the hierarchical complexity of topologies. Firstly, a new metric to measure hierarchical complexity; secondly, a Weighted Complex Hierarchy (WCH) model. To thoroughly evaluate our techniques, we generalise sparse binary network archetypes to weighted forms and explore the main topological features of brain networks- integration, regularity and modularity- using curves over density. 

\textit{Results}: By controlling the parameters of our model, the highest complexity is found to arise between a random topology and a strict 'class-based' topology. Further, the model has equivalent complexity to EEG phase-lag networks at peak performance.

\textit{Comparison to existing methods}: Hierarchical complexity attains greater magnitude and range of differences between different networks than the previous commonly used complexity metric and our WCH model offers a much broader range of network topology than the standard scale-free and small-world models at a full range of densities.

\textit{Conclusions}: Our metric and model provide a rigorous characterisation of hierarchical complexity. Importantly, our framework shows a scale of complexity arising between 'all nodes are equal' topologies at one extreme and 'strict class-based' topologies at the other.

\end{abstract}


\end{frontmatter}

\section{Introduction}

Graph theory is an important tool in functional connectivity research for understanding the interdependent activity occurring over multivariate brain signals \cite{BullmoreSporns2009, Stam2014b, Papo2014}. In this setting, Complete Weighted Networks (CWNs) are produced from all common recording platforms including the Electroencephalogram (EEG), the Magnetoencephalogram (MEG) and functional Magnetic Resonance Imaging (fMRI), where every pair of nodes in the network share a connection whose weight is the output of some connectivity measure. Complex hierarchical structures are known to exist in real networks \cite{Ravasz2003}, including brain networks \cite{BullmoreSporns2009, Meunier2010}, for this reason it is important to find methods to specifically evaluate hierarchical complexity of network topology. Here we introduce methods specific to this end.

Complexity is understood neither to mean regularity, where obvious patterns and repetition are evident, nor randomness, where no pattern or repetition can be established, but attributed to systems in which patterns are irregular and unpredictable such as in many real world phenomena \cite{Costa2005}. Particularly, the brain is noted to be such a complex system \cite{Tononi1994} and this is partly attributed to its hierarchical structure \cite{Meunier2010}. Hierarchical complexity is thus concerned with understanding how the hierarchy of the system contributes to its complexity. Here we introduce a new metric aptly named hierarchical complexity, $R$, which is based on targeting the structural consistency at each hierarchical level of network topology. We compare our metric with network entropy \cite{Sole2004} and find that we can offer a greater magnitude and density range for establishing differences in complexity of different graph topologies.


Alongside this, we introduce the Weighted Complex Hierarchy (WCH) model which simulates hierarchical structures of weighted networks. This model works by modifying uniform random weights by addition of multiples of a constant, which is essentially a weighted preferential selection method with a highly unpredictable component provided by the original random weights. We show that it follows very similar topological characteristics of networks formed from EEG phase-lag connectivity. Intrinsic to our model is a strict control of weight ranges for hierarchical levels which offers unprecedented ease, flexibility and rigour for topological comparisons in applied settings and for simulations in technical exploration for brain network analysis. This also provides an unconvoluted alternative to methods which randomise connections \cite{WattsStrogatz1998,Sporns2006} or weights \cite{RubinovSporns2011} of the original network.

Any rigorous evaluation of brain networks should address their inherent complete weighted formulation \cite{Fallani2014}. However, the current field has largely lacked any concerted effort to build an analytical framework specifically targeted at CWNs, preferring instead to manipulate the functional connectivity CWNs into sparse binary form (e.g.  \cite{Sporns2006,Li2011a,Tewarie2015} as well as wide-spread use of the Watts-Strogatz \cite{WattsStrogatz1998} and Albert-Barabasi \cite{AlbertBarabasi1999} models) and using the pre-existing framework built around other research areas which have different aims and strategies in mind \cite{Newman2010}. In our methodological approach we propose novel generalisations of pre-existing sparse binary models to CWN form and thus allow a full density range comparison of our techniques. Due to the intrinsic properties of these graph types we find minimal and maximal topologies which can help to shed light on a wide variety of topological forms and their possible limitations \cite{Sole2004} in a dense weighted framework.

Further, as part of our study we seek after straightforward metrics to evaluate other main aspects of network topology for comparisons  \cite{Sole2004,Sporns2010}  and, in this search, found it necessary to revise key network concepts of integration-segregation \cite{Stam2014b,WattsStrogatz1998,RubinovSporns2009} and scale-freeness \cite{AlbertBarabasi1999,Eguiluz2005}. We provide here these revisions: i) That the clustering coefficient, $C$, is enough to analyse the scale of integration and segregation, finding it unnecessary and convoluted to use the characteristic path length, $L$, as a measure of its opposite, as generally accepted \cite{BullmoreSporns2009,WattsStrogatz1998}. ii) We provide mathematical justification that the degree variance, $V$, and thus network irregularity \cite{Snijders1981} is a strong indicator of the scale-free factor of a topology.

Our study of hierarchical complexity, using a comprehensive methodological approach, provides mathematical quantification of the hierarchical complexity of EEG functional connectivity networks and reveals new insights into key aspects of network topology in general. Our model provides improved comparative abilities for future clinical and technical research.

\section{Network Science: Proposed methods and key revisions}
We adopt the notation in \cite{SandryhailaMoura2013} so that a graph, $G(\mathcal{V},\textbf{W})$, is a set of $n$ nodes, $\mathcal{V}$, connected according to an $n\times n$ weighted adjacency matrix, $\mathbf{W}$. Entry $W_{ij}$ of $\mathbf{W}$ corresponds to the weight of the connection from node $i$ to node $j$ and can be zero. An unweighted graph is one in which connections are distinguished only by their existence or non-existence, so that, without loss of generality, all existing connections have weight $1$ and non-existent connections have weight $0$. The graph is undirected if connections are symmetric, which gives symmetric $\mathbf{W}$. A simple graph is unweighted, undirected, with no connections from a node to itself and with no more than one connection between any pair of nodes. This corresponds to a graph with a symmetric binary adjacency matrix with zero diagonal. Such graphs are easy to study and measure \cite{Newman2010}. The degree, $k_{i}$, of node $i$ is defined as the number of its adjacent connections, which is the number of non zero entries of the $i$th column of $\mathbf{W}$. Then, for a simple graph, $k_{i} = \sum_{j=1}^{n}W_{ij}$. For a graph with $2m$ edges, the connection density, $P$, of a graph is $P = 2m/n(n-1)$. 

A CWN is represented by a symmetric adjacency matrix with zero diagonal (no self-loops) and weights, $W_{ij}\in[0,1]$, elsewhere. To analyse CWNs it is beneficial to convert it to simple form by binarising the adjacency matrix using a threshold, where a percentage of strongest connections are set to $1$ and the remaining values set to $0$. This stays true to the network activity \cite{Fallani2014} whilst reducing computational complexity and weight issues found with weighted metrics \cite{Stam2014b}. Hereafter, all mathematics will refer to simple graphs.

In this section we present the contributions of this study. We first present the hierarchical complexity metric and the WCH model, which are the key novel contributions of this paper. Thereafter we detail revisions and clarification of integration and segregation as a scale evaluated by $C$ and scale-freeness as a factor evaluated by $V$. Finally, we outline the generalisation of key network archetypes to CWN form, full details of which can be found in the supplementary material.

\subsection{Hierarchical Complexity Metric}
The ideas of order and complexity are well known in the discussion of networks (indeed, real world networks are often called complex networks \cite{BullmoreSporns2009,Papo2014,McAuley2007}). In mathematics, the graphs studied derive from some theoretical principles. These can involve set patterns, without random fluctuations of connections, such as regular networks, fractal networks, star networks and grid networks. On the other hand much interest is shown in more randomly generated topologies, such as random graphs and other graphs involving random processes, as these express something of the more erratic and irregular quality of connections in networks constructed from real world phenomena \cite{WattsStrogatz1998,Erdos1959}. However, real world phenomena differ from random processes in that there is a clear organisational behaviour apparent throughout the hierarchical structure, both within hierarchical levels and between hierarchical levels \cite{Ravasz2003,Meunier2010}. Although this structure is perhaps impossible to retrace, because its formation inevitably involves many unknown generative processes, we can provide methods for its analysis.

\begin{figure}[!t]
	\centering
	\includegraphics[trim= 0 25 0 0,clip,width= 3.5in]{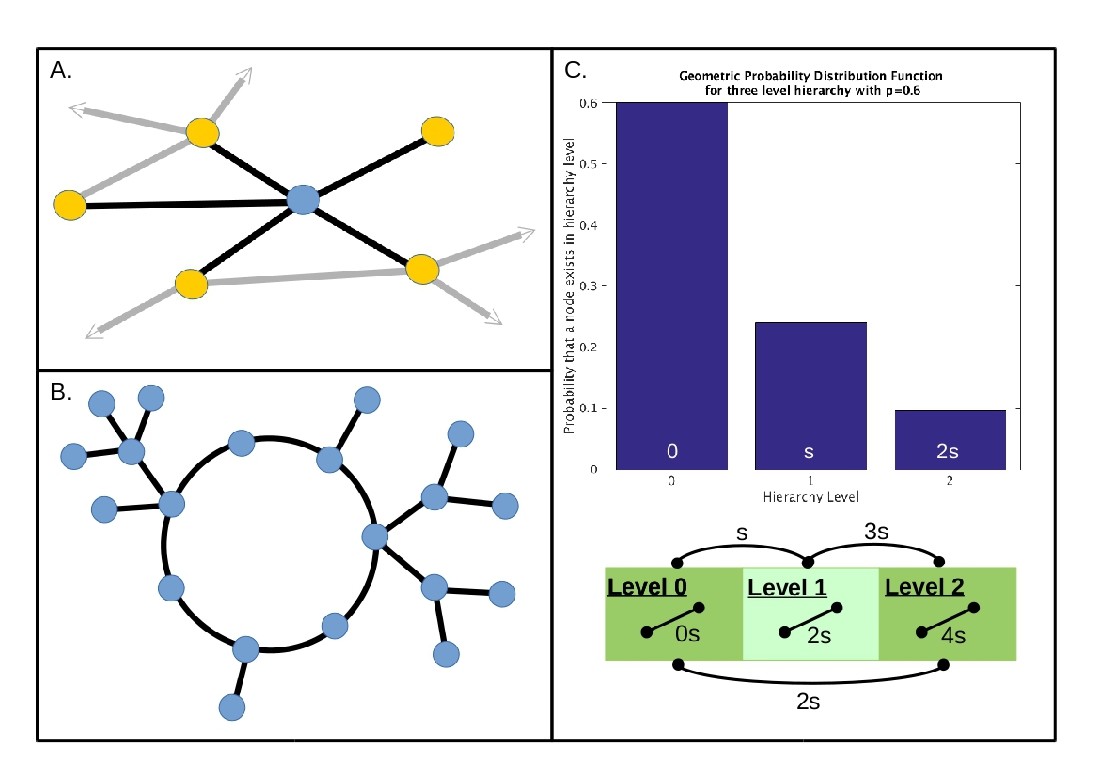}
	\caption{A. Example of a node degree neighbourhood. Here is shown a part of a network relating to the neighbourhood of the blue node. The blue node has neighbourhood degree sequence $\{1,2,3,4,4\}$, i.e. the ordered degrees of the orange nodes. Grey connections indicate all the additional connections of the orange nodes in the network. B. Example for graph complexity. Here is shown a 20 node network with varying 'orderedness' at different degree levels. C. Diagram of the construction of the WCH model. Above is the probability distribution function for a geometric distribution with $p = 0.6$ for a three level hierarchy. Below is a graphic displaying the additional weight added between nodes in given hierarchy levels.}
	\label{fig1}
\end{figure}

Hierarchies in networks are generally determined by degrees of nodes, where a small group of highly connected nodes create a rich club \cite{McAuley2007} on the top hierarchical level and nodes with generally lesser connectivity exist on a peripheral lower levels. Further, it is seen that a node's relationship within the context of the network is greatly determined by the other nodes to which it is connected \cite{Bona2001}. Thus, to understand the hierarchical complexity of a network we propose to study the behaviour of nodes of a given degree by looking at the degrees of nodes in their neighbourhoods. We define $D$ as the set of degrees of a graph, $G$. Similar to the idea of node degree sequences \cite{Molloy1995}, we can construct neighbourhood degree sequences, specific to each node in the graph. That is, for a node, $i$, of degree $k \in D$ we have a sequence 
\[
s_{i} = \{d_{i,1}, d_{i,2}, …, d_{i,k}\}\ \text{ s.t. }  d_{i,1} \leq d_{i,2} \leq\dots\leq d_{i,k} \in D, 
\]
where $d_{i,j}$ is the degree of the $j$th node connected to node $i$ (see Fig. \ref{fig1}.A). For all nodes of a given degree, $k$, the corresponding neighbourhood degree sequences have equal length, $k$.

We define the hierarchical complexity, $R$, of a network as the average variance of the $k$-degree neighbourhood degree sequences and can be expressed as:
\begin{equation}
R=\frac{1}{\mathcal{D}}\sum_{D_{k}\neq \emptyset}\frac{1}{kr_{k}(r_{k}-1)}\left(\sum_{j=1}^{k}\left(\sum_{i\in D_{k}}(s_{ki}(j)- \mu_{kj})^{2}\right)\right),
\end{equation}
where $\mathcal{D}$ is the number of distinct degrees in the graph, $D_{k}$ is the set of nodes of degree $k$, $s_{ki}(j)$ is the $j$th element of the $i$th $k$-length sequence, $\mu_{kj}$ is the mean value of element $j$ over all $k$-length sequences and $r_{k}$ is the number of nodes of degree $k$, which is added to the denominator for normalisation of hierarchy levels.

Organisation of the graph at the level of $k$-degree nodes can be seen by comparing the $j$th elements of their neighbourhood sequences. If all of the $j$th elements of all the sequences are equal, that is $s_{i}=s_{j}$ for all $s_{i}$, $s_{j}$ of length $k$, then there is a high degree of order present in the $k$-degree nodes of the graph. If these sequences differ widely however, then it can be said that the $k$-degree nodes are either disorganised or more complexly organised. For example, in Fig.\ref{fig1}.B the two degree nodes all have the same degree sequences- $\{3,4\}$- whereas the three degree nodes are split into two different degree sequences- $\{1,2,2\}$ and $\{1,1,4\}$- and finally the neighbourhood degree sequences of the four degree nodes are all different- $\{1,1,1,4\}$, $\{1,2,2,4\}$ and $\{2,3,3,3\}$. So the complexity of just the two degree nodes is 0, the complexity of just the three degree nodes is $2((2-1.5)^{2}+(1-1.5)^{2} + (2-3)^{2} + (4-3)^{2})/(4\times 3\times 3) = 5/36$ and the complexity of just the four degree nodes is $(2(1-4/3)^{2}+(2-4/3)^{2} + 2((1-2)^{2} + (3-2)^{2}) + 2(4-11/3)^{2}+(3-11/3)^{2})/(4\times 3\times 2) = (16/3)/24 = 8/36$, the complexity over all three levels being the average- 13/108.

This measure is thus minimal for graphs in which, for each $k$ and $k'$, every $k$-degree node is connected to exactly the same number of $k'$-degree nodes. This property, for example, is seen in ring lattices, and quasi-star graphs and is close to minimal in the line graph, fractal graphs and grid lattices. Furthermore, the degrees of random networks are known to have a fairly small spread which is a factor penalised by our complexity value. Thus random networks should obtain low values of our complexity measure. On the other hand, $R$ values of real networks are expected to be higher given the high spread and degree fluctuations of those networks caused by hub nodes promoting a high degree irregularity while the spontaneous nature of real-world connections should promote a high variability of the neighbourhood degree sequences.

\subsection{Weighted Complex Hierarchy Model} 
The foundation of our model is the random CWN model. The most general random network is the Erd\"os-R\'enyi (E-R) random network \cite{Erdos1959} which is formed by assigning a probability, $p$, to the question of the existence or non-existence of connections on a network with $n$ nodes. Such a construct is, in fact, an ensemble of graphs denoted $G(n,p)$. A sample of this ensemble is obtained by generating a random value for every possible connection and applying the probability value $p$ as a threshold to see whether or not that connection should exist in our sample. The random CWN model is thus simply a symmetric matrix with zero diagonal and randomly generated values $W_{ij}\in[0,1]$ elsewhere. If we threshold the CWN at weight $T = p$, we recover a binary Erd\"os-R\'enyi random graph from the random graph ensemble $G(n,p)$.

Starting from an Erd\"os-R\'enyi CWN we randomly distribute the nodes into hierarchy levels based on some discrete cumulative distribution function, $\mathbf{p}$, by generating a random number, $r$, between 0 and 1 for each node and putting the node in the level for which $r - \mathbf{p}$ is first less than $0$. We then distribute $ls$ additional weight to all connections of adjacent nodes in the $l$th level, for some suitably chosen $s$. The parameters of this model are then $(n,s,l,\mathbf{p})$. The parameter $n$ is the number of nodes in the network. The parameter $s$ is the strength parameter, which is constant since the random generation of the initial weights is enough to contribute to weight randomness. The parameter $l$ is the number of levels of the hierarchy, with a default setting of a random integer between $2$ and $5$. The vector $\mathbf{p}$ is the cumulative probability distribution vector denoting the probabilities that a given node will belong to a given level where the default, which we use here, is a geometric distribution with $p=0.6$ in hierarchical levels ($0,1,2,\dots,l$) where the nodes with highest connectivity (top hierarchical level) are at the tail end of the distribution. Fig. \ref{fig1}.C plots an example of the geometric distribution for a three level hierarchy. The text inside the box plots, above, indicates the additional weights given to connections adjacent to nodes inside the given level. The graphic below explains the additional weights provided by the strength parameter of connections between nodes in different levels as well as in the same level. For example, a connection between a level 1 node and a level 2 node has additional strength $3s$ which consists of one $s$ provided by the node in Level 1 and $2s$ provided by the node in Level 2. At $s = 0$, we have the E-R random network and at $s = 1$ the weights of the network are linearly separable by the hierarchical structure producing a strict 'class-based' topology. Between these values a spontaneous 'class-influenced' topology emerges.

\subsection{Revision of concepts from network science}
Here we present justifications for metrics as measures of key topological factors- the global clustering coefficient, $C$, for degree of segregation and the degree variance, $V$, for irregularity, linked to scale-freeness.
\subsubsection{Integration-Segregation}
The concept of integration in brain networks is closely tied in to the small world phenomenon \cite{Milgram1967}, where real world networks are found to have an efficient 'trade off' between segregative and integrative behaviours \cite{BullmoreSporns2012}. The most widely used topological metrics in network science- $C$ and the characteristic path length, $L$- are commonly noted as measures of these quantities, respectively. Here, $L$ is defined as the average of the shortest paths between each pair of nodes and $C$ is defined as the probability that a path of length $2$, or triple, in the graph has a shortest path of length $1$. That is, 
\begin{equation}
C  = \frac{\text{closed triples}}{\text{triples}},
\end{equation}
where a closed triples is such that, for triple $\{W_{ik}, W_{kj}\}$, $W_{ij} = 1$, for  $i,j,k$ distinct.


Since integration implies a non-discriminative behaviour in choice, we argue that the random graph ensemble  \cite{Erdos1959}, defined by its equal probability of existent
connections between all pairs of nodes, is the most exemplary model of an integrated network. Anything which deviates from equal probability is a discriminative factor which favours certain connections or nodes over others, likely leading to more segregated activity. Further, it is clear that integration and segregation are opposite ends of the same spectrum- something which is not integrated must be segregated and vice versa. Having one metric to inform on where a network lies on that spectrum is therefore sufficient. 

Thus, here we propose $C$ as the topological measure to evaluate levels of integration (and so segregation) of a given network. Firstly, we note that values of $C$ for random graphs and small-world graphs are often much more distinguishable than those of $L$ \cite{WattsStrogatz1998} and it is certainly assumed that these graphs have very different levels of integration. Secondly, since the random network is optimally integrated and $E[C_{\text{ran}}] = E[P_{\text{ran}}]$ \cite{Newman2010}, where $P_{\text{ran}}$ is the connection density of the random network, then the larger the deviation from $1$ of the value $\gamma = C/E[C_{\text{ran}}] = C/E[P_{\text{ran}}] = C/P$, the more segregated is the network. We will include both $L$ and $C$ in our analysis in order to provide evidence to back the above proposal.

\subsubsection{Regularity and Scale-Freeness}
Another topological factor of small world networks is noted as a scale-free nature characterised by a power law degree distribution \cite{Barabasi1999}. To understand this aspect of network topology another factor of network behaviour is formulated distinguishing between 'line' like and 'star' like graphs \cite{Stam2014b,Tijms2013}. 

Here, we show that characterisation of scale-freeness is closely connected to the regularity of a network. Regular graphs have been studied for over a century \cite{Petersen1891}. They are defined as graphs for which every node has the same degree. An almost regular graph is a graph for which the highest and lowest degree differs by only $1$. Thus a highly irregular graph can be thought of as any graph whose vertices have a high variability. Such behaviour can be captured simply by the variance of the degrees present in the graph, that is
\begin{equation}
V =\text{var}(D),
\end{equation} 
where $D = \{k_{i}\}_{i\in\mathcal{V}}$, is the set of node degrees on a given graph \cite{Snijders1981}. 

For regular graphs $V=0$ by definition, but more probing is necessary to distinguish high $V$ topology. For a graph with degrees $\mathbf{k} = \{k_{1},k_{2},\dots,k_{n}\}$, and $\sum_{i=1}^{n}k_{i}=2m$, on multiplying out the brackets $V$ simplifies to 
\begin{align*}
V &= \frac{1}{(n-1)} \sum_{i=1}^{n}\left(\frac{2m}{n} -  k_{i}\right)^{2}\\ 
  &= \frac{\|\mathbf{k}\|_{2}^{2}}{(n-1)}  - 2mP,  
\end{align*}
where $P = 2m/n(n-1)$ is the connection density and $\|\textbf{k}\|_{2}^{2}=\sum_{i=1}^{n}k_{i}^{2}$, is the squared $\ell_{2}$ norm of $\mathbf{k}$. This tells us that $V$ is proportional to the sum of the squares of the degrees of the graph, $\|\textbf{k}\|_{2}^{2}$, and, for fixed number of connections, $m$, $V$ in fact depends only on $\|\textbf{k}\|_{2}^{2}$. Now, it is known that $\|\textbf{k}\|_{2}^{2}$ is maximal in quasi-star graphs and quasi-complete graphs \cite{Abrego2008}. Essentially, the quasi-star graph has a maximal number of maximum degree nodes in the graph for the given connection density and the quasi-complete graph has a maximal number of isolated, or zero-degree, nodes in the graph. This tells us that, for low $P$, high $V$ denotes the presence of a few high degree nodes and a majority of relatively low degree nodes, i.e. scale-free-like graphs. Thus, due to the restriction placed on possible degree distributions by the number of edges (the small number of edges in sparse networks means the number of high degree nodes is very limited), the irregularity of degrees is a strong indicator of the strength of 'decay' of the given distribution, relating to how 'scale-free' the graph is.

\subsection{Complete Weighted Network Archetypes}
\begin{figure}[!t]
	\centering
	\includegraphics[trim= 50 550 300 50,clip, width = 2.8in]{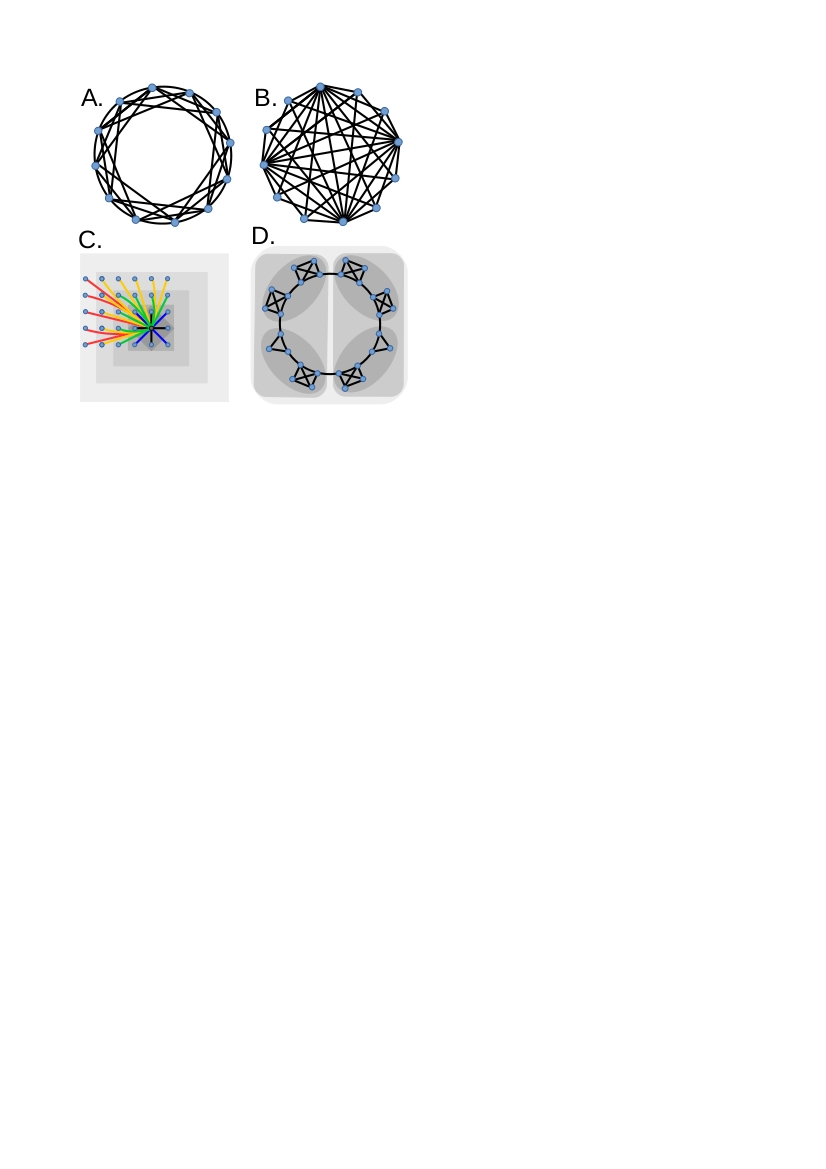}
	\caption{A. A 12 node ring lattice of degree $6$, comprising the three strongest weight categories of the ring lattice CWN. B. The quasi-star with $4$ nodes of degree $n-1$ and $n-4$ nodes of degree $4$, also comprising the first four categories of the star CWN. C. The grid lattice weight categorisation (relating to the grey node) in a $30$ node network (see supplementary material). Colours of edges denote category: black, blue, green, orange and red edges are in weight categories $1$, $2$, $3$, $4$ \& $5$, respectively. The increasingly lighter boundaries thus represent 'catchment' areas around the node by increasing category. Centring these 'catchment' areas around a given node gives the respective categorisation of edges adjacent to the new node. D. Fractal modular CWN weight categorisation on 30 nodes. Edges shown (black) are $1$st weight category edges. In this instance, increasingly lighter background represents areas within which all pairs of nodes become connected by edges when the network is subject to the threshold corresponding to the respectively increasing category (see supplementary material).}
	\label{ordered}
\end{figure}

In the supplementary material we detail the method to generalise sparse binary network archetypes to CWN form. The pre-requisit of such a generalisation is that we require obvious higher density versions of lower density forms which can be arranged in adjacency matrix form such that each non-zero entry, $W_{ij}$, of the lower density adjacency matrix exists as a non-zero entry, $\hat W_{ij}$ in the higher density adjacency matrix. This is indeed the case for the Regular Ring Lattice, Star, Grid Lattice and Fractal Modular CWNs (see Fig. \ref{ordered} A,B,C,D respectively). We explain these higher and lower density forms of the binarised CWN in terms of weight categories where, if we choose an appropriate threshold, $T$, we can recover all edges in the same and all higher weight categories and none of the edges existing in all lower categories.

\section{Methods}

Here we apply methods to graphs of $64$ nodes, typical of medium density EEG. For analysis we employ connection density thresholds at integer percentages of strongest weighted connections, rounded to the nearest whole number of connections. We then implement metric algorithms on each of these binary networks and plot the obtained values on a curve against connection density, similar as in e.g. \cite{Stam2007,Lynall2010}. This generates metric curves plotted against connection density which provides a detailed analysis of the CWN topology. Other methods exist to analyse CWNs such as weighted metrics \cite{RubinovSporns2011} or density integrated metrics \cite{Ginestet2011}, but these metrics still give only singular values for a given network which belies little of topological behaviour at different scales of connectivity strength.

For random and WCH CWNs we use sample sizes of $100$ for each network and for the EEG functional connectivity CWNs we have a sample size of $109$ \cite{Goldberger2000}. On the metric curves for these we plot the median with the interquartile range shaded in. For ordered networks there is only one network per type by definition.

Our analytical framework is composed of a mixture of entirely new concepts and novel generalisations of existing concepts to CWN form. It is constituted of the following elements:
four metrics, $R$, $C$, $V$, $Q$  characterising four important and distinct topological features; five CWN archetypal models- Random, Star, Regular Lattice, Fractal Modular, Grid Lattice; the  WCH model.

\subsection{Metrics}
In \cite{Sole2004,Sporns2010} an 'architecture' of network topology is proposed involving the three most widely studied properties of brain networks- integration (and segregation) \cite{Stam2014b,WattsStrogatz1998,RubinovSporns2009}, 'scale-freeness' \cite{AlbertBarabasi1999,Eguiluz2005} and modularity \cite{Meunier2010,NewmanGirvan2004}. For our analysis in comparison with hierarchical complexity, $R$, we choose a straightforward metric for each of these topological factors- $C$ for integration, $V$ for scale-freeness and $Q$ for modularity \cite{NewmanGirvan2004} where
\begin{equation}
Q	= 	\frac{1}{2m} \sum_{i,j}\left(W_{ij} - \frac{k_{i}k_{j}}{2m}\right) \delta(c_{i},c_{j}),
\end{equation}
where $c_{i}$ is the module containing node $i$ and $\delta()$ is the Kronecker delta function. Highly efficient algorithms have been created \cite{Newman2006,Blondel2008} aiming to maximise the value of $Q$ for a given network. To compute the modularity of our networks, we use the undirected modularity function \cite{Newman2006} in the Brain Connectivity Toolbox \cite{RubinovSporns2009}.

\subsection{Comparison for Hierarchical Complexity}
We compare our hierarchical complexity metric with a commonly used metric for analysing the entropy of the network degrees \cite{Sole2004}. This is defined using the normalised degree distribution $q_{i} =  k_{i}p_{i}/\sum_{j}k_{j}p_{j}$, where $k_{i}$ is the degree of node $i$ and $p_{i}$ is the proportion of nodes in the graph with the same degree as node $i$ which relates to probabilities of ‘going to’/ ‘coming from’ neighbouring nodes in directed graph problems. Then the entropy of graph $G$ is a straightforward derivation of Shannon's entropy equation  \cite{Shannon} for the degrees of the graph:

\begin{equation}
H(G) = -\sum_{i=1}^{n} q_{i} \log(q_{i}).
\end{equation}

Thus, Network entropy encodes the eccentricity of the graph degrees.

\subsection{Comparisons for the WCH model}
We implement comparisons with the Watts-Strogatz small-world model \cite{WattsStrogatz1998} which randomly rewires a set proportion of edges starting from a regular lattice. We use the full range of parameters for initial degree specification (2 up to 62) and random rewiring parameters from 0.05 in steps of 0.05 up to 0.95. For each combination of parameters, 100 realisations of the model were computed and $C$, $V$, $Q$, and $R$ were measured. We further compare with Albert-Barabasi's scale-free model \cite{AlbertBarabasi1999} which begins with a graph consisting of core of highly connected nodes to which the rest of the nodes are added one by one with a set degree but paired by edges to randomly selected nodes. We use an initial number of nodes of 15 and the additional node's degree from 3 up to 14 in order to reach larger densities.

\subsection{EEG networks} 
We use an eyes open, resting EEG data set with $64$ nodes. We report on networks created from the beta ($12.5$-$32$Hz) band using coherence and the debiased Weighted Phase-Lag Index (dWPLI) in order to account for different possible types of EEG networks while reducing redundancy of similar topological forms found between the frequency bands (see supplementary material).

The dataset, recorded using the BCI2000 instrumentation system \cite{Schalk2004}, was freely acquired from Physionet \cite{Goldberger2000}. The signals were recorded from 64 electrodes placed in the main in accordance with the international 10-10 system. We took the eyes open resting state condition data, consisting of $1$ minute of continuously streamed data which were partitioned into $1$s epochs and averaged for each of $109$ volunteers.

FieldTrip \cite{Oostenveld2011} was used for pre-processing, frequency analysis and connectivity analysis to obtain the adjacency matrices of complete weighted networks. The $64$ channels were re-referenced using an average reference, the multi-taper method was implemented from $0$ seconds onwards using Slepian sequences and $2$Hz spectral smoothing. A $0.5$Hz resolution was obtained using one second of zero padding. We chose to analyse the matrices obtained from both the coherence and the debiased Weighted Phase-Lag Index (dWPLI) \cite{Vinck2011} to look for differences between network topologies of zero and non-zero phase lag dependencies in the channels \cite{vanDiessen2014}. We treat the data of all tasks as a single dataset to allow for the variability of the EEG network topologies since we are not interested here in the tasks themselves but on the behaviour of general EEG networks obtained from dWPLI and coherence. 

\subsection{Statistical analysis}
Due to the polynomial formulation of the complexity measure, producing a non-normal distribution, we compare metric distributions using the Wilcoxon rank sum test. The z-score is used to ascertain the magnitude and direction of the relationship of the distributions.

\section{Results}
\subsection{Metric Comparisons}
\begin{figure}[!t]
	\centering 
	\includegraphics[trim= 40 40 0 0,clip,width = 3.8in]{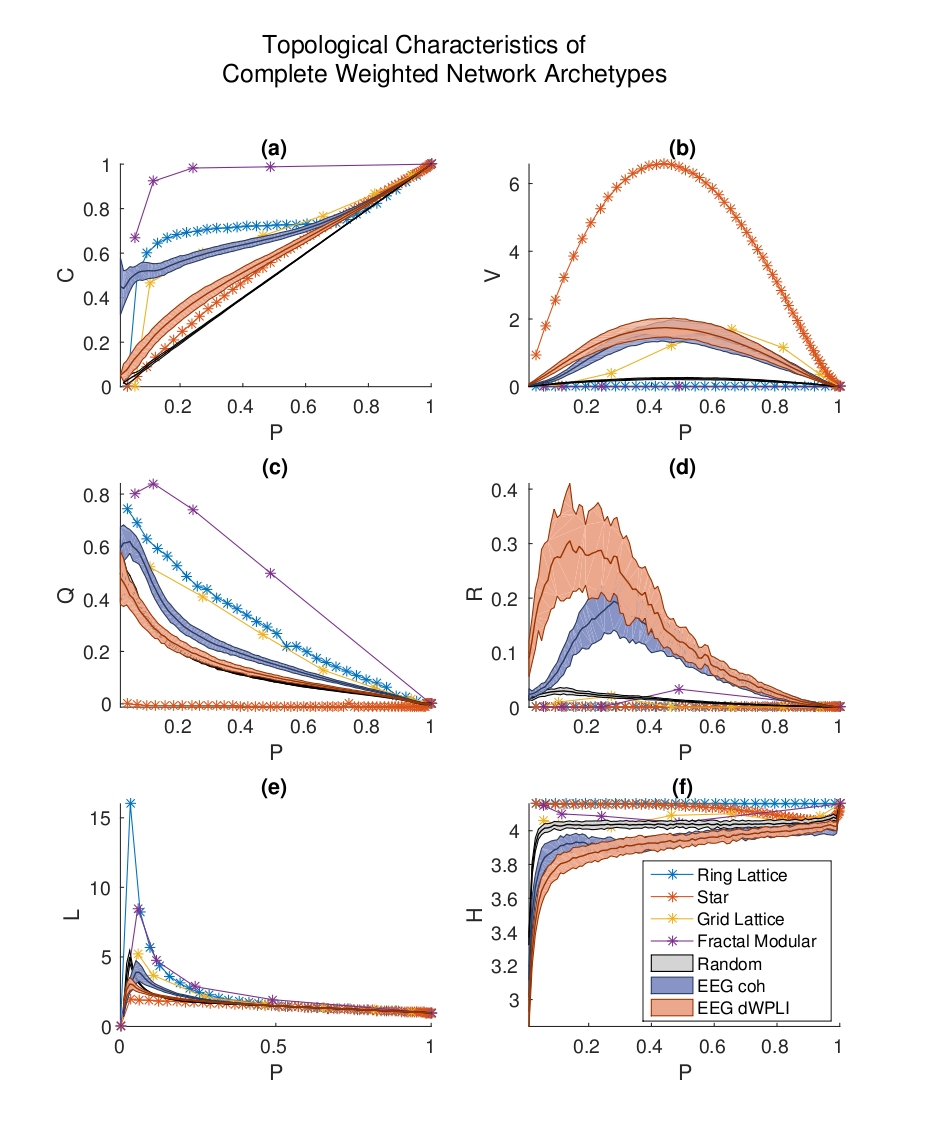}
	\caption{Topological metric values for integration ($C$), regularity ($V$), modularity ($Q$), hierarchical complexity ($R$), characteristic path length ($L$) and network entropy ($H$) against network density, $P$. Curves relate to network models as indicated in the legend (bottom right).}
	\label{fig3}
\end{figure}

Fig. \ref{fig3} shows the metric curves (i.e. metric plotted against network density) for $C$, $V$, $Q$, $R$, $L$ and $H$ for all archetypes as well as for the EEG dWPLI (red shade) and coherence (blue shade) networks. From these plots we see experimental evidence of maximal and minimal topologies for the given topological characteristics. These maximal and minimal topologies are explained as the curves whose lines are consistently lowest or highest over all densities. Fractal Modular networks (purple lines) are maximal for both $C$ and $Q$ (top left and centre left, respectively). This is to be expected since the modules are complete sub-networks with very few connections between modules, maximising $Q$. Further this minimises the number of open triples in the graph, maximising $C$, by restricting open triples to relating only to those few connections which do extend between modules. The star CWN (orange lines) acts as a maximal topology for $V$, as expected from the theory explained in Section 2, while being a minimal topology for $L$ (bottom left). Regular graphs, such as the ring lattice network (blue lines), give $0$ degree variance and hierarchical complexity, thus are minimal topologies of these features. The results of Fig. \ref{fig3} for 30 node networks, found in the supplementary material, follow the same relationships, providing evidence that these features are independent of network size.

Comparing the plots in Fig. \ref{fig3} of $C$ (top left) with $L$ (bottom left) and $R$ (centre right) with $H$ (bottom right), it is immediately clear that $L$ and $H$ show extreme behaviour at low densities while remaining consistent at higher densities. This exemplifies how these metrics are aimed at analysis of sparse networks, where it appears that values can take a much greater range than for higher density networks.

\begin{figure}[!t]
	\centering 
	\includegraphics[trim = 120 0 0 0,clip,width=3.8in]{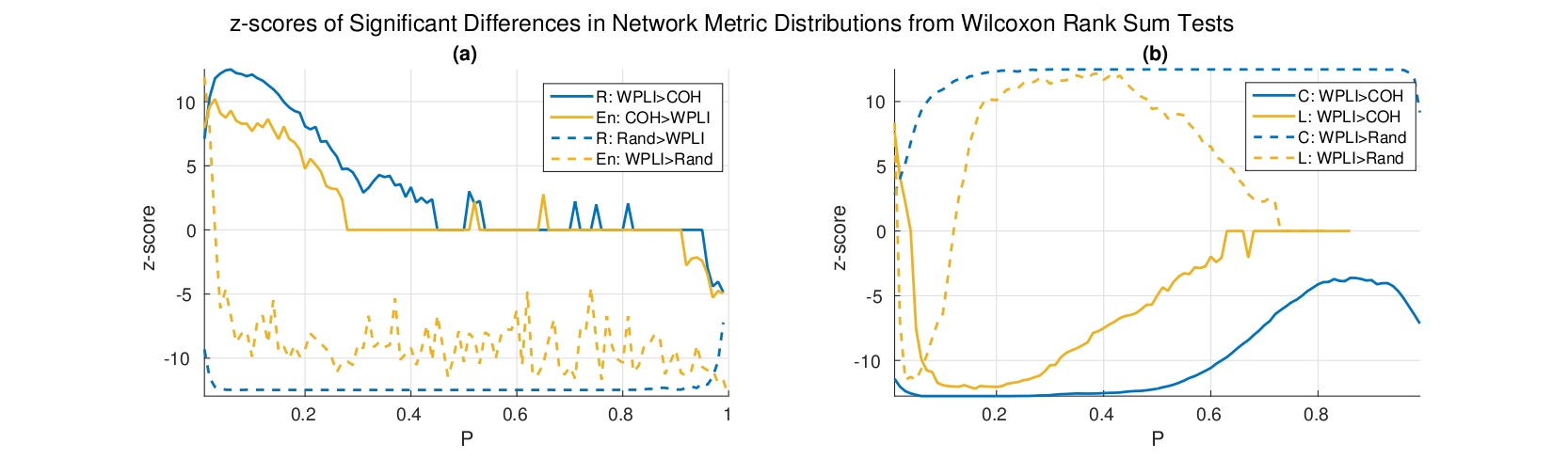}
	\caption{Positive (negative) values indicate the contrasted distributions exhibit the relationship provided in the legend (or its opposite). Zero indicates $p$-value insignificant at 5\% level. a) The hierarchical complexity, $R$ (blue), compared with network entropy, $H$ (yellow). b) The clustering coefficient , $C$ (blue), compared with characteristic path length, $L$ (yellow). $P$ is the network density.
	}
	\label{CompareMetrics}
\end{figure}

To explore these comparisons further we perform statistical analysis with Wilcoxon rank sum tests on the differences of distributions of metric values of EEG dWPLI and E-R random networks as well as of EEG dWPLI and EEG coherence networks (Fig \ref{CompareMetrics}). The results show that $C$ (right) and $R$ (left) attain a greater range over edge density, $P$, of significant differences than their counterparts, $L$ and $H$. Particularly, $R$ distinguishes differences from 1\% up to 44\% densities in the EEG dWPLI and coherence comparison (solid blue line), whilst entropy only can distinguish differences from 1\% up to 27\% (solid yellow line). Further, the z-scores indicate that in the range 1-27\%, the differences found in $R$ are greater than those found using $H$. Comparing the EEG dWPLI networks with E-R random networks (Fig \ref{CompareMetrics}, left, dashed lines), both metrics find differences at all levels, but the magnitude of difference found by $R$ (blue) is consistently greater than those found by $H$ (yellow). Thus, our metric outperforms entropy in both magnitude and range of differences found. 

Similarly, $C$ finds a greater range and magnitude of differences than $L$, Fig \ref{CompareMetrics} right. In fact, $C$ discerns differences at all connection densities for the two comparisons, while $L$ fails to find differences after 62\% in comparing dWPLI and coherence networks (solid yellow line) and after 73\% in comparing dWPLI and random networks dashed yellow line). Furthermore, $L$ displays inverse differences at low densities (1-12\%) compared to higher densities in the dWPLI vs random comparison (dashed yellow line). This inconsistency is undesirable for translatability of integrative behaviour of network types from sparse networks to more dense networks. $C$ does not suffer from such behaviour, displaying a constant relationship of metric values through the full range of densities (solid and dashed blue lines).

Given these results, for the rest of our analysis, we will drop $L$ and $H$ and focus on the four proposed metric, $C$, $V$, $Q$ and $R$. We must emphasise that this is taken purely in terms of the simplicity of explaining a general topological factor and does not mean that $L$ and $H$ are not useful for other purposes.

\subsection{Weighted Complex Hierarchy Null Model}

\begin{figure}[!t]
	\centering 
	\includegraphics[trim = 40 20 0 0,clip,width=3.8in]{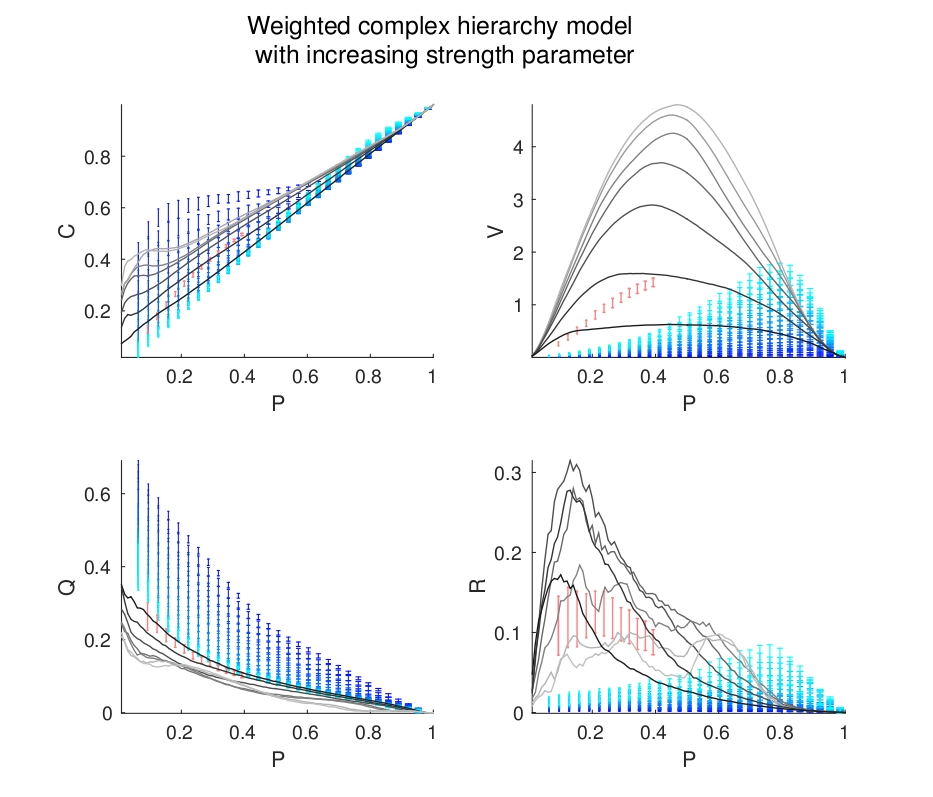}
	\caption{The topological characterisation of network models by clustering coefficient, $C$, degree variance, $V$, modularity, $Q$ and hierarchical complexity, $R$, plotted against network density, $P$. Grey lines indicate mean values of the weighted complex hierarchy model with increasing light shade indicating increasing strength parameter from $s = 0.1$ in steps of 0.1 up to $s = 0.7$. Red errorbars indicate values of the Albert Barabasi scale-free model. Blue errorbars indicate values of the Watts Strogatz small-world model with increasingly light blue indicating increasing proportion of edges being randomly rewired.}
	\label{WCHModel}
\end{figure}

Fig. \ref{WCHModel} shows the mean results of $C$ (top left), $V$ (top right), $Q$ (bottom left) and $R$ (bottom right) over 100 realisations of each of the WCH models. We include a reduced number of strength parameters in the figure ($s = 0.1, 0.2,\dots, 0.7$) than those computed ($s = 0.05, 0.1,\dots, 0.75$) for greater clarity. Above 0.75 the parameter begins to saturate as the weights of the hierarchy levels tend to linear separability (linear separability occurs when $s=1$ since $0s,1s,2s,...$ then places the edge weights, originally in $[0,1]$, in disjoint ranges $[0,1],[1,2],[2,3],...$). We see that WCH networks (grey shaded lines) exhibit curve behaviour similar to the EEG networks and E-R random graphs (as in Fig. \ref{fig3}). The scale-free model (red error bars) also exhibits a similar behaviour, however in stark contrast, the small-world model (blue error bars) exhibits very different behaviours than those of the EEG or WCH networks, exhibiting a strong unsuitability for comparisons with EEG networks with much higher modularity and highly right skewed $V$ curve (Fig. \ref{WCHModel}, top left) towards high densities as well as a similar right skew in $R$ (bottom left) which is opposite to the left skew found for WCH and EEG network types. Although the scale-free model exhibits similar tendencies in topological metrics to the WCH and EEG networks, its range of values and densities is clearly very limited and so, therefore, its ability for topological refinement.

By increasing the strength parameter of the WCH model we change the topology in a smooth fashion with decreasing integration, regularity and modularity (Fig. \ref{WCHModel}, top left, top right and bottom left, respectively). Interestingly, $R$ (bottom left) rises with increasing strength parameter from $s = 0.05$ up to $s = 0.3$ where it takes its maximum values at densities ranging from 1-30\% before falling again from $s = 0.35$ until $s = 0.7$. Further, above $s = 0.3$, the curves begin to deviate significantly from those of the EEG dWPLI networks, exhibiting greater plateaus of high complexity (lighter grey lines) which are more comparable with the EEG coherence networks.

\begin{figure}[!t]
	\centering 
	\includegraphics[trim = 0 0 0 0,width=3in]{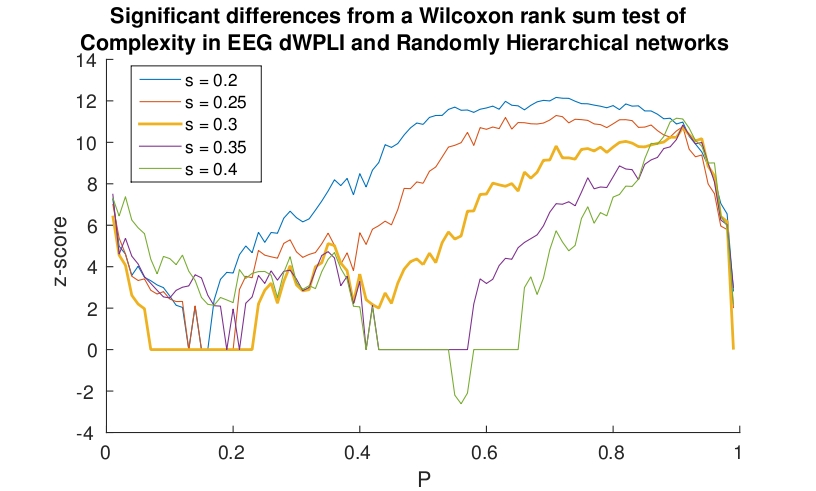}
	\caption{The z-statistics of distributions with significant differences
		from a Wilcoxon rank sum test. Positive (negative) values indicate the contrasted distributions exhibit the relationship provided in the legend (or its opposite). Zero indicates $p$-value insignificant at 5\% level. $P$ is the connection density.}
	\label{CompareModel}
\end{figure}

Interestingly, the complexity of the EEG dWPLI networks appears to attain maximal values of $R$ of all the networks studied here (Fig. \ref{fig3}). The only model which comes close is the WCH model (Fig. \ref{Mimic}, bottom right). To clarify this observation we perform Wilcoxon rank sum tests on $R$ values of the EEG dWPLI networks against that of the WCH model with strength parameters ranging from $s = 0.2$ up to $s = 0.4$, i.e. two steps before and after the maximal complexity setting of $s = 0.3$. The results are displayed in Fig. \ref{CompareModel}. In the vast majority of instances of strength parameter and density, the EEG dWPLI networks do indeed exhibit greater complexity than the WCH model. The strong exception to this is an inability to distinguish significant differences between the maximal complexity $s = 0.3$ WCH model and dWPLI networks within 7-23\% densities (bold yellow line). Also, as the weight parameter increases, the high plateaus previously mentioned begin to take effect as in the medium ranges of density the $R$ values of the dWPLI networks and WCH model becomes more indistinguishable, with greater complexity found in the range 55-57\% in the WCH model with $s = 0.4$ (green line). 

\subsection{Null model approaching EEG phase-lag networks}

\begin{figure}[!t]
	\centering 
	\includegraphics[trim = 40 20 0 0,clip,width=3.8in]{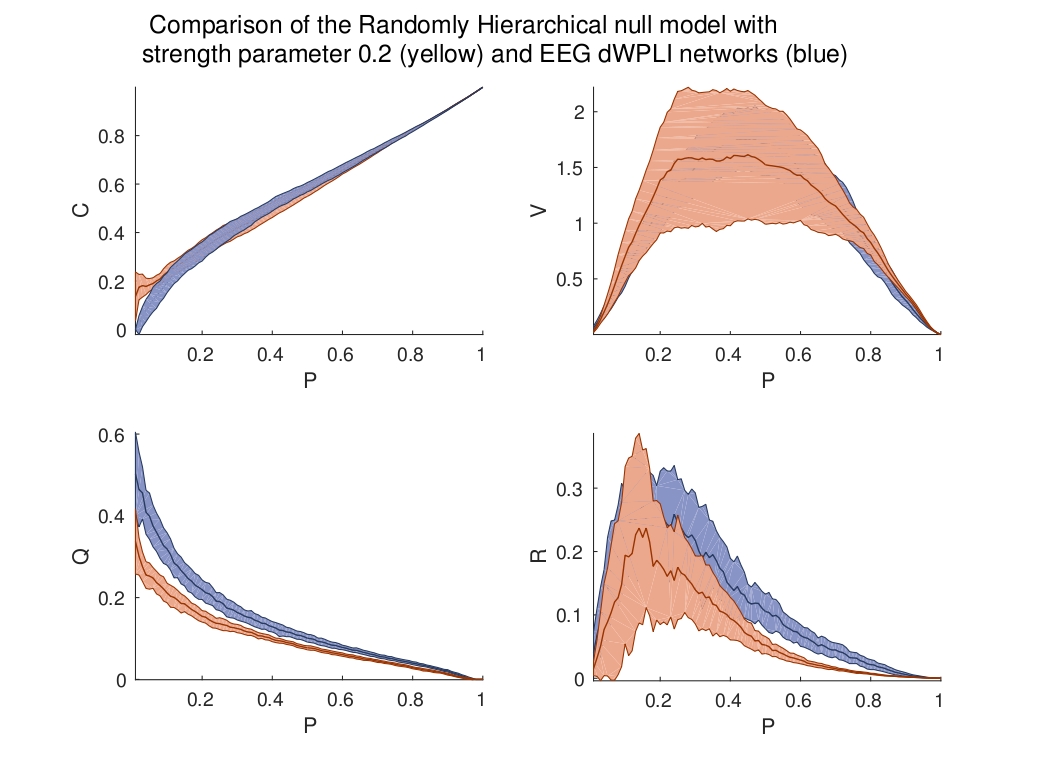}
	\caption{Clustering coefficient, $C$, degree variance, $V$, modularity, $Q$, and complexity, $R$, against connection density, $P$, of binarised weighted networks, for WCH model (red, median (m) $\pm$ interquartile range (iqr)) and EEG dWPLI neworks (blue, m $\pm$ iqr).}
	\label{Mimic}
\end{figure}

Fig. \ref{Mimic} shows the values of the four topological features- complexity, integration, regularity and modularity for EEG dWPLI networks and the WCH network with strength parameter $0.2$. We see clearly that these networks behave very similarly with respect to the given metrics. The most obvious difference is that the modularity, $Q$, of dWPLI EEG networks is higher (bottom left). Also, as previously discussed, the dWPLI network complexity is greater than the WCH model, but it is still by far the most comparable model for complexity of those presented here.

\section{Discussion}

\subsection{Complexity as revealed by weighted complex hierarchy model}
The behaviour demonstrated by the WCH model with respect to $R$ indicates that high complexity arises from a hierarchical structure in which a greater degree of variability is present in the rankings of weights with respect to hierarchy level. Too little difference between levels and the hierarchy is too weak to maintain complex interactions, too much difference between levels and the complexity of the hierarchy is dampened by a more ordered structure produced from the tendency towards linear separability of the edge weights enforced by the strength parameter. Thus, we provide evidence that topological complexity is not driven by integration, arising as a middle ground between regular and random systems as previously conjectured \cite{Tononi1994,WattsStrogatz1998}, but, driven by hierarchical complexity, arising in the middle ground between weak hierarchical topology or 'all nodes are equal' systems, such as random or regular networks, and strong hierarchical topology, such as star or strict class-based systems including grid lattice and fractal modular networks (see Fig. \ref{WCHModel}). Thus the hierarchical structure can be seen as a key aspect of the complexity inherent in complex systems.

Impressively, the dWPLI EEG networks display a generally greater hierarchical complexity than that expressed by our model which is specifically designed to probe complex interactions in hierarchical structures. Thus we pose such complexity as a key aspect of brain function as modelled by phase-based connectivity.

\subsection{Weighted complex hierarchy as null model}
There are two clear reasons why the WCH model is a good fit for functional connectivity networks from EEG recordings. Not only does it create several hub like nodes giving a high degree variability, but furthermore it simulates the rich club phenomena found in complex brain networks \cite{McAuley2007,Heuvel2011}, as the higher the hierarchy levels of two nodes, the stronger the weight of the connection will be between them, see Fig. \ref{fig1}.C.

One of the greatest benefits of this model over others is that it simulates brain networks previous to network processing steps because it creates CWNs rather than sparse networks. This means that any and all techniques one wants to use on the brain networks can be applied elegantly and in parallel with this single null model free from any complications. Particularly, methods which create sparse binary networks directly, whether these models are built independently from the brain networks \cite{Sporns2006,AlbertBarabasi1999} or are constructed by the randomisation of connections of the networks being compared \cite{WattsStrogatz1998,RubinovSporns2009}, run into problems with density specification (in the case of independent models) and reproducibility (in both types of model). With the WCH model, we can simply create a bank of simulated CWNs which can be used throughout the study in exactly the same way as we use the functional connectivity CWNs.

As an example of the power and elegance of the proposed model, say we want to find maximum spanning trees  \cite{Stam2014a} of our brain networks and compare with a null model, then we simply take the maximum spanning trees of our null model. In contrast, in \cite{Tewarie2015} they use a convoluted reverse engineering process by assigning random weights to the connections of Watts-Strogatz small world networks (which are themselves of limited comparability to brain networks) and computing the MST from these resulting sparse weighted networks.

Further, as seen in Fig. \ref{fig1}.C, for technical studies which rely on network simulations, the WCH model is built on parameters which can be altered to subtly change the resulting topology. This allows for sensitive analysis of a new techniques ability to distinguish subtle topological differences. Such paradigms are evident in clinical studies where, for example, one may try to distinguish between healthy and ill patients \cite{Tijms2013,Lynall2010} or between different cognitive tasks \cite{Smith2015}, so that this null model offers simulations which are directly relatable to clinical settings.

\subsection{EEG coherence and WPLI networks}
We see there is a large difference in the integration, modularity and complexity of the EEG coherence and dWPLI networks (Fig. \ref{fig3}, top left, centre left and centre right, respectively). The EEG coherence networks (blue shade) behave similarly to the ring (blue lines) and grid lattice (yellow lines) networks, agreeing with the volume conduction effects that dominate zero-lag dependency measures \cite{vanDiessen2014}, i.e. the closer the nodes are the stronger the weights are. The dWPLI networks (red shade) on the other hand have a more integrated and less modular nature, which reflects the notion that phase-based functionality mitigates volume conduction effects and is thus less confined by anatomical structure \cite{vanDiessen2014}.

The very high complexity of the dWPLI networks (and very possibly phase-lag measures in general \cite{Dauwels2010}) provides evidence to support that phase-based connectivity does indeed largely overcome the volume conduction effect and therefore maintains a richer complexity echoing the complex interactions of brain functionality \cite{BullmoreSporns2009}.

With regards to how the WCH model advances our understanding of dWPLI and coherence network differences, we note that the high segregation of the coherence networks (Fig. \ref{fig3} top left) is approached by the WCH model with high values of strength parameter (Fig. \ref{WCHModel}, top left) and is comparable with regular lattice and grid lattice CWN curves (Fig. \ref{fig3}, top left, blue and yellow lines, respectively), denoting a move to a more strict class-based topology. This is also reflected in the hierarchical complexity (bottom right of corresponding figures), where the lower complexity peaking at a later density to dWPLI (Fig. \ref{fig3}, centre right) is mimicked in the behaviour of increasing strength parameter in the WCH model (Fig. \ref{WCHModel}, bottom right). This provides further evidence of the relevance and flexibility of the WCH model. In contrast there is an evident lack of ability to make similar comments with respect to the popular small world and scale-free models. This criticism can be extended towards network models which randomise connections while maintaining degree distributions \cite{RubinovSporns2009}, since such an enforced topological attribute does not allow one to analyse how that very important attribute is actually constructed. Future work will provide extension to modular structures in our model to focus on what roles modularity plays on these aspects, since $Q$ and $V$ behave contrastingly to this extrapolation.

\subsection{Dense scale-free networks}
A striking feature seen is in the degree variance curves where a highly symmetric parabolic curve is noted with a central maximum value for random graphs, WCH networks and EEG networks. This feature reveals to us a 'scale-free' paradigm at all density levels and not just the classic sparse network scale-free at low densities. In other words, the scale-free nature found in brain networks is first and foremost encoded in the connectivity weights, which, through selective binarisation, therefore can reveal to us the scale-free property as expressed at different density ranges. As the density of the network increases one obtains more even distributions of high and low density nodes, indicated by the high values of $V$, and, eventually, towards high densities the symmetry of $V$ values with low densities tells us that the scale-free network is characterised by a small number of low degree nodes and a majority of high degree nodes, i.e. the inverse (or complement) of the low density behaviour.
	
\subsection{Topological randomness}
If we define a uniformly random topology as that which exhibits a uniform distribution of topological values over the space enveloped by the minimal and maximal topologies, it is very apparent that E-R random networks do not satisfy this criteria, but, instead, have a restricted topology at all density levels where the interquartile range is much smaller in comparison with that of the EEG networks and the proposed null model. We thus see that uniformly distributed random weights do not lead to a uniformly random topology in this sense, but instead to a very particular optimally integrated, moderately regular, lowly modular and low complexity topology at all densities. Based on this evidence and previous discussion of random networks in the methods section, we suggest that E-R random networks should be re-understood as optimally integrated networks.

Following from this the randomisation of connections used widely in null models is not a topologically randomising process but, more accurately, a topologically integrative process. Such a feature is then not necessarily typical of network topology and thus one must be cautious to use this as a null model unless one wants to specifically target integrative behaviour. Further, the practice of normalisation of graph values by E-R random graph values \cite{HumphriesGurney2008} must also be used with due caution. The basis of such a normalisation is to contrast a networks values with those of the 'average' network topology \cite{Bol1998}, rather than contrasting with a highly specific topology which behaves very differently to real world networks \cite{Newm2005}. This evidence provides further justification for the adoption of our WCH model as a relevant and powerful replacement to these models.

\section{Conclusion}
We introduced a metric for measuring the hierarchical complexity of a network and a highly flexible and elegant WCH model. These provided key insights into what distinguishes functional brain networks from both ordered and spontaneous forms as generally the most complex kind of topology and the important role that hierarchical structure plays in this. Further, we showed that phase-based connectivity topology was more complex than amplitude influenced connectivity topology, which we extrapolated as due to the more ordered structure enforced by volume conduction effects. In our analysis we constructed a framework for CWNs for brain functional connectivity to replace the framework for sparse networks adopted from other network science research areas. This included the synthesis of concepts from the literature in a succinct manner and the generalisation of sparse binary archetypes to CWN form. The perspective allowed by this comprehensive analysis provided new evidence regarding key factors of network topology in general. Importantly we provided evidence of the non-topologically random nature of uniformly random weighted networks. From this it follows that our model is more relevant and appropriate than prevalent connection randomisation processes. Also, a scale-free paradigm was extended to all network densities. Particularly, these insights help towards a comprehensive understanding of the framework within which functional connectivity networks are set and thus provide invaluable information and tools for future clinical and technical research in neuroscience. Matlab codes for all synthesis and analysis of the networks as introduced in this paper are publicly available on publication at http://dx.doi.org/10.7488/ds/1520.


\section{Acknowledgements}
Keith Smith is funded by the Engineering and Physical Sciences Research Council (EPSRC).

\end{document}